\documentclass[twocolumn,nofootinbib,preprintnumbers,amsmath,amssymb,superscriptaddress]{revtex4}%prb
\usepackage{epsfig}
\pdfoutput=1

\usepackage[mathletters]{ucs}
\usepackage[utf8x]{inputenc}

\usepackage{amsmath,amssymb,amsthm,amsfonts,mathtools}
\usepackage{mathrsfs}
\usepackage{slashed}
\usepackage{graphicx}
\usepackage{subfigure}
\usepackage[dvipsnames]{xcolor}
\usepackage{rotating}
\usepackage{dsfont}
\usepackage{setspace}
\usepackage{verbatim}
\usepackage{fancyhdr}
\usepackage
[pdftitle={Generative Neural Samplers for the Quantum Heisenberg Chain},
pdfsubject={Generative Neural Samplers for the Quantum Heisenberg Chain}]
		{hyperref}
\usepackage{cleveref}
\usepackage{tabularx}
\usepackage{algorithm,algpseudocode}
\usepackage{siunitx}

\newcommand{\ktj}{kT/J}
\graphicspath{{./figures/}}

\DeclareMathOperator{\Tr}{Tr}
\DeclareMathOperator{\csch}{csch}
\DeclareMathOperator{\sech}{sech}
\DeclareMathOperator{\E}{\mathbb{E}}
\newcommand{\ham}{\mathcal{H}}
\newcommand{\obs}{\mathcal{O}}
\newcommand{\bra}[1]{\ensuremath{\left\langle#1\right|}}
\newcommand{\ket}[1]{\ensuremath{\left|#1\right\rangle}}

\begin{document}
\title{Generative Neural Samplers for the Quantum Heisenberg Chain} 

\author{Johanna Vielhaben}
\email{johanna.vielhaben@hhi.fraunhofer.de}
\affiliation{Fraunhofer Heinrich Hertz Institute, 10587 Berlin, Germany}

\author{Nils Strodthoff}
\email{nils.strodthoff@hhi.fraunhofer.de}
\affiliation{Fraunhofer Heinrich Hertz Institute, 10587 Berlin, Germany}

\begin{abstract}
Generative neural samplers offer a complementary approach to Monte Carlo methods for problems in statistical physics and quantum field theory.
This work tests the ability of generative neural samplers to estimate observables for real-world low-dimensional spin systems. 
It maps out how autoregressive models can sample configurations of a quantum Heisenberg chain via a classical approximation 
based on the Suzuki-Trotter transformation. We present results for energy, specific heat and susceptibility for the isotropic XXX and 
the anisotropic XY chain that are in good agreement with Monte Carlo results within the same approximation scheme.
\end{abstract}

\maketitle

\section{Introduction}
Monte Carlo methods are the most important numerical techniques for gaining insights in statistical mechanics systems and lattice field theories from first principles
\cite{mc_newman, gattringer_lang}. Despite their success, Monte Carlo methods come with some drawbacks. They often suffer from long auto-correlation times, especially close to criticality. Moreover, they can get trapped in a subset of configuration space which hinders them from sampling all relevant parts of the space. 

Generative neural samplers (GNSs) are deep learning models that can sample from an explicitly or implicitly learned probability distribution. They were successfully used to model complicated real-world distributions in applications like text-to-speech or image synthesis. Recently, they have been applied to a variety of problems in statistical physics and lattice field theory. Some approaches rely on samples from a Monte Carlo algorithm for training while others are completely independent from Monte Carlo methods. GNS models of the first-mentioned category have been used for example to estimate the thermodynamics of the Ising model in two dimensions \cite{thermodynamics_BM, DL_Ising, GAN_ising, ising_field_VAE}. Representatives of the latter category are autoregressive models or normalizing flows and offer a truly complementary approach to Monte Carlo methods. In \cite{wu_solving_stat_mech}, autoregressive models were applied to the two-dimensional classical Ising and Sherrington-Kirkpatrick model. The approach was advanced in \cite{nicoli2019comment,kim_savan} by bringing forward an asymptotically unbiased estimator for GNSs that provide the exact sampling probability.

In the domain of quantum systems, GNS have been used for the study of ground-state properties, offering an alternative approach to methods like Variational Monte Carlo or Green's function Monte Carlo. In \cite{BM_Carleo, BM_Carleo_2, BM_noe, BM_groundstate1, DNN_groundstate}, GNS have shown the ability to represent the ground-state of quantum many-body-systems. Moreover, GNS have found application in the domain of lattice field theories, like for the $\varphi^4$-theory \cite{zhou2019regressive,phi4_urban,phi4_shanahan,nicoli2020estimation} and for the $U(1)$-gauge theory \cite{lft_u1} and $SU(N)$-gauge theory \cite{lft_SUN} in two dimensions.

In this work, the journey towards autoregressive models for the thermodynamics of quantum mechanical statistical models is taken. Here, the quantum Heisenberg chain in two special cases serves as an example. This system is approximately realized in real materials \cite{Ising_realization, XY_realization, XXZ_realization, realization_QPT} and is a key model in studying quantum phase transitions or the dynamics of correlated lattice systems \cite{QPT_1, QPT_2,santos_dynamics}.

The paper is organized as follows: \Cref{sec:heisenberg_gns} gives a short overview of the quantum Heisenberg chain, maps out the Suzuki-Trotter transformation, which is used here to approximate the quantum Heisenberg chain by a classical system, and explains how samples from the classical approximation can be drawn by a GNS model. \Cref{sec:results} examines the estimates for internal energy, specific heat and susceptibility of the isotropic and XY Heisenberg chain. We summarize and conclude in \Cref{sec:conclusions}. Technical details and derivations were included in several appendices to make the paper self-contained.

\section{Generative neural samplers for the Heisenberg chain}
\label{sec:heisenberg_gns}
\subsection{Heisenberg chain}
\label{sec:heisenberg}
The general spin-$\frac{1}{2}$ Heisenberg chain with nearest-neighbor coupling and without an external field is described by the Hamiltonian
\begin{equation} \label{eq:qhc_hamiltonian}
	\ham = -\sum_{i=1}^N \left( J_x S_i^x S_{i+1}^x  +  J_y S_i^y S_{i+1}^y  +  J_z S_i^z S_{i+1}^z \right)\,.
\end{equation}
Here, $S_i^{x,y,z}$ are the spin operators that act on each site $i$ and fulfill the commutation relations $[S_i^a,S_j^b]=\delta_{ij}\epsilon^{abc}S_i^c$. $J_x, J_y, J_z$ are real-valued coupling constants and we assume periodic boundary conditions by identifying $S_{N+1}^{x,y,z}\equiv S_1^{x,y,z}$. The spin-$\frac{1}{2}$ Heisenberg chain is solved exactly by the Bethe Ansatz in the sense that the spectrum of the Hamiltonian is described exactly by the Bethe equations \cite{bethe_original}. Because of the increasing computational complexity of identifying all Bethe solutions when $N \rightarrow \infty$ this knowledge about the spectrum of the Hamiltonian does not enable a direct quantification of the thermodynamic behavior. To do so, numerical calculations for finite-size systems can be used and extrapolated to infinite chains, like in the pioneering work of \cite{calc_bonnerfisher_64} with chain lengths up to $11$. Two well-known methods to derive the thermodynamic quantities of the XXX chain are the thermal Bethe Ansatz developed by \cite{tba_takahashi} and the Quantum Transfer Matrix Method pioneered by \cite{QMT_start_koma, QMT_start_suzuki}.

In this work, we study two special cases: The ferromagnetic XXX Heisenberg chain with $J_x=J_y=J_z>0$ and the ferromagnetic XX-chain with $J_x=J_y>0$ and $J_z=0$, a special case of the XY-chain. In the case of the XY-chain, where the coupling between the $z$-spin components is turned off in \Cref{eq:qhc_hamiltonian}, an analytical solution of the free energy is available. It was first obtained by \cite{xy_LSM}, who applied a Jordan-Wigner transformation under which the XY chain reduces to a system of free spinless fermions. Shortly after, \cite{xysol_katsura_62} used a different procedure to write the Hamiltonian of the XY chain with a transverse external field in terms of fermion creation and annihilation operators. The resulting expression for the free energy of a chain of $N$ spins with periodic boundary conditions in the limit $N \rightarrow \infty$ and without a magnetic field is
\begin{align} \label{eq:xyF}
	F =& - \frac{NkT}{2\pi} \int_0^{\pi}  \ln \Bigl[ 2\cosh \bigl( K_x^2 +K_y^2 \notag \\
	  &+2K_xK_y \cos(2\omega)\bigr) \Bigr]^{1/2}\,d\omega\,,
\end{align}
where $K_x=\frac{J_x}{4kT}$ and $K_y=\frac{J_y}{4kT}$.
The GNS results of this work will mainly be compared to results from a Markov Chain Monte Carlo (MCMC) algorithm by \cite{mc_cullenlandau_83}, which is described in \Cref{app:mcmc}. It is based on the mapping of the quantum spin chain to a classical two-dimensional system discussed below and a special update routine.

\subsection{Suzuki-Trotter transformation} \label{sec:suzuki_trotter}
\begin{figure}[htpb]
	\centering
	\includegraphics[width=\columnwidth]{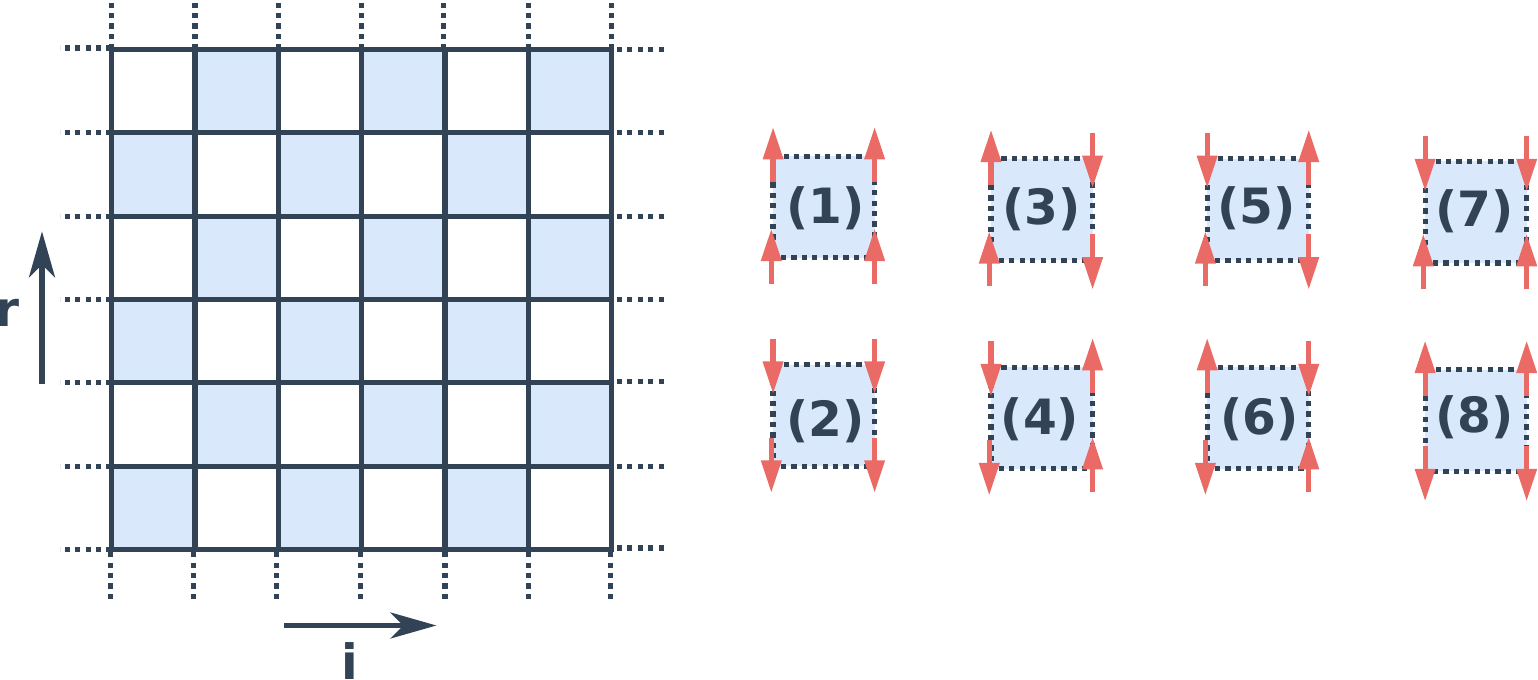}
	\caption{Left: The transformed Heisenberg model is a 2-dimensional Ising model with 4-spin interactions represented by blue squares. Right: Allowed 4-spin interactions }
	\label{fig:plaquettes}
\end{figure}

In order to apply GNSs to the quantum Heisenberg chain, the Suzuki-Trotter decomposition is used to transform the partition function of the quantum spin system,
\begin{equation}
	Z = \Tr  e^{-\beta \ham} \,,
\end{equation} 
into the partition function of a two-dimensional Ising-like classical system \cite{checkerboard_barma_78, mc_cullenlandau_83},
\begin{multline} \label{eq:4spin_ising}
	Z^{(m)} = \sum_{\alpha_1 \alpha_2 \dots \alpha_{2m}} \exp \Bigl( 2K_0 \sum^N_{i=1}\sum_{r=1}^{2m}S_{i,r}S_{i+1,r}  \\  - \beta \sum_{\hat i}\sum_{\hat r} h(i,r) \Bigr) \,,
\end{multline}
with $K_0 = \frac{\beta}{4m}J_z$. The caret denotes that the sum is running over terms with $i$ and $r$ both odd or both even and $S_{i,r}=\pm\frac{1}{2}$ are eigenvalues of the Ising part of $\ham$ in \Cref{eq:qhc_hamiltonian}. The matrix elements $h(i,r)$ couple four spins and are found in \Cref{eq:4spin_matrixel}. The derivation of \Cref{eq:4spin_ising} is mapped out in \Cref{sec:st_derivation}. \Cref{eq:4spin_ising} can be interpreted as the partition function of an $N\times2m$ Ising system with two-spin interactions along the real space direction $i$ and temperature-dependent four-spin interactions on the alternating plaquettes marked blue in \Cref{fig:plaquettes} that couple neighboring spins in both the real direction and the Trotter direction $r$. When evaluating the matrix elements in \Cref{eq:4spin_matrixel} one finds that only $8$ of the $16$ possible cases for a 4-spin plaquette are non-zero, namely those with an even number of spins in each direction. Only configurations that contain just the non-zero 4-spin plaquettes are allowed, as configurations violating these constraints would have infinite energy or zero contribution to the partition function otherwise.  
From \Cref{eq:trace_operation} one can read of that the trace operation in $Z^{(m)}$ requires periodic boundary conditions in the Trotter direction. Following \cite{mc_cullenlandau_83}, periodic boundary conditions are also used in the real space direction.
The eight allowed 4-spin plaquettes are depicted in \Cref{fig:plaquettes}. Their energies, including the 2-spin interaction from the first term and the 4-spin interaction from the second term in \Cref{eq:4spin_ising}, are

\begin{align}
	E^{(m)}(1,2) &= -\frac{1}{\beta} \left( K_0 + \ln \cosh K_- \right) \label{eq:E12} \\
	E^{(m)}(3,4) &= \frac{1}{\beta} \left( K_0 - \ln \cosh K_+ \right) \label{eq:E34}\\
	E^{(m)}(5,6) &= \frac{1}{\beta} \left( K_0 - \ln \sinh K_+ \right) \label{eq:E56} \\
	E^{(m)}(7,8) &= -\frac{1}{\beta} \left( K_0 + \ln \sinh K_- \right) \label{eq:E78}
\end{align}

When $J_x\to J_y$, $K_-\to 0$ and hence $E^{(m)}(7,8)$ diverges. Thus, only configurations that contain the 4-spin plaquettes $(1)$-$(6)$ are allowed in this case.

\subsection{Thermal averages} \label{sec:thermal_av}
The thermal properties of the classical system above can be found by taking the respective derivatives of the free energy. The resulting expressions differ from the naive ones due to the temperature-dependent interactions in the effective Hamiltonian \cite{mc_cullenlandau_83}. The thermal average of the \textit{internal energy} is
\begin{align}
	E^{(m)} &= \frac{1}{Z^{(m)}} \sum_j F_j^{(m)} \exp \left( -\beta E_j^{(m)} \right)\,,
\end{align}
where $F_j^{(m)}$ is the non-trivial \textit{energy function} in the $j$th state whose ensemble average is the thermal average, i.e.\
\begin{equation}
	E^{(m)} = \langle F^{(m)} \rangle\,.
\end{equation}
From there, the \textit{specific heat} can be derived,
\begin{equation}
	C^{(m)} = \frac{1}{T^2} \left( \langle F^{(m)}{}^2 - G^{(m)} \rangle - \langle F^{(m)} \rangle^2 \right)\,,
\end{equation}
with
\begin{equation}
	G_j^{(m)} = \frac{\partial^2}{\partial \beta^2} \left( \beta E_j^{(m)} \right)\,.
\end{equation}
The \textit{parallel susceptibility} is obtained as 
\begin{equation}
	\chi^{(m)} = \beta \left( \langle M^{(m)^2} \rangle - \langle M^{(m)} \rangle^2 \right)\,,
\end{equation}
where 
\begin{equation}
	M_j^{(m)} = \frac{1}{2m} \left( \sum_i\sum_r S_{i,r} \right)\,.
\end{equation}
A detailed derivation of these thermal averages and the contributions of an individual plaquette to $F_j^{(m)}$, $G_j^{(m)}$ and $M_j^{(m)}$ are given in \Cref{sec:thermal_av_derivation}.

\subsection{Autoregressive models}
This work focuses on autoregressive models which form a subclass of GNSs. They provide the exact probability $q_\theta(s)$ with which a configuration $s \sim q_\theta$ is sampled, where $\theta$ parameterizes the sampler. To model $q_\theta(s)$ for an $N$-dimensional distribution $p(s)$ an ordering $s_1, \dots, s_N$ of the components of $s$ is specified. The one-dimensional conditional distributions $q_\theta(s_i|s_{i-1}, \dots, s_1)$ are modeled by a neural network.
The joint probability is the product of the conditional distributions over all components,
\begin{equation} 
	q_\theta(s) = \prod_{i=1}^N q_\theta(s_i|s_{i-1}, \dots, s_1)\,, \label{eq:jointprob}
\end{equation}
and a sample can be drawn from $q_\theta$ by sequentially sampling from the corresponding conditional distributions.
Autoregressive models can be trained by minimizing the inverse Kullback-Leibler divergence between the Boltzmann distribution and the variational distribution $q$,
\begin{align} \label{eq:KL_div}
	D_\text{KL}(q_{\theta} |p ) &= \sum_s q_{\theta}(s) \ln \left( \frac{q_{\theta}(s)}{p(s)} \right) 	 \notag\\
				  &= \sum_s q_{\theta}(s)  \left( \ln \left( q_{\theta}(s) \right) + \beta \ham (s) \right)  + \ln(Z)\,,
\end{align}
without resorting to Monte-Carlo configurations. The expression can be straightforwardly optimized by gradient descent upon noting the that last term, $\ln(Z)$, only represents an irrelevant constant in this respect. In this work, we use a PixelCNN \cite{vdoord_pcnn} to model the conditional distributions $q_\theta(s_i|s_{i-1}, \dots, s_1)$, which was developed in the context of natural image generation. In \cite{wu_solving_stat_mech, nicoli2019comment, kim_savan} it was applied to classical two-dimensional spin systems.
\subsection{Sampling configurations}
The mapped Heisenberg model only allows the 4-spin configurations depicted in \Cref{fig:plaquettes}. When a autoregressive model samples a configuration spin-by-spin and row-by-row, the outcome will in general include illegal plaquettes. In early tests, a high energy was assigned to these configurations as a penalty term to see whether the model would learn to avoid sampling illegal configurations. This led to a quick collapse of the training and the model would learn to only sample either spins up ($S=+1$) or spins down ($S=-1$). The strategy used instead was to change the sampling procedure such that only configurations with exclusively legal plaquettes are generated. In this procedure, the spins are sampled row-by-row, but only if the value of a spin is not determined by the condition to generate only legal plaquettes. For a plaquette in the center for example, only the first three spins of a plaquette are sampled. The fourth spin is determined by the product of the first three spins as the product of all four spins is required to be positive. When using this modified sampling procedure, only the conditional probabilities that were used to sample a spin contribute to the joint probability and \Cref{eq:jointprob} changes to
\begin{equation}
	q_\theta(s) = \prod_{\substack{\text{sampled} \\ \text{positions}\;i}} q_\theta(s_i|s_{i-1},\dots,s_1)
\end{equation}
\cite{mc_cullenlandau_83} applied their Monte Carlo method to systems with $J_x=J_y$ where plaquettes (7) and (8) are forbidden. It is not possible to construct a sequential sampling procedure that produces configurations with only the first six plaquettes from \Cref{fig:plaquettes} that obeys the periodic boundary conditions in both directions. To circumvent this problem, very similar systems with $J_x=J_y+\epsilon$ with $\epsilon=$\SI{1e-6}\;are studied. For the XY chain, the analytical solutions for the internal energy $E$, specific heat $C$ and susceptibility $\chi$ by Katsura \cite{xysol_katsura_62} in \Cref{eq:xyE,eq:xyC,eq:xyChi} show that there is basically no difference between $\epsilon=$\SI{1e-6} and $\epsilon=0$.

\section{Results}
\label{sec:results}
Following \cite{mc_cullenlandau_83}, Heisenberg chains of length $L=32$ were investigated, where finite size effects are small compared to statistical errors in the Monte Carlo data. The GNS samplers were tested for $m=2$ and $m=4$, corresponding to approximate classical systems on $4 \times 32$ and $8 \times 32$ grids, respectively. Next to the simple sample mean in \Cref{eq:simple_mean}, the asymptotically unbiased Neural Importance Sampling (NIS) estimator in \Cref{eq:NISestimator} by \cite{nicoli2019comment, kim_savan} is evaluated for each observable. The experimental details are listed in \Cref{app:exp_det}.
\subsection{XY chain}
\label{sec:XYchain}
\begin{figure}[htpb]
	\centering
	\includegraphics[width=\columnwidth]{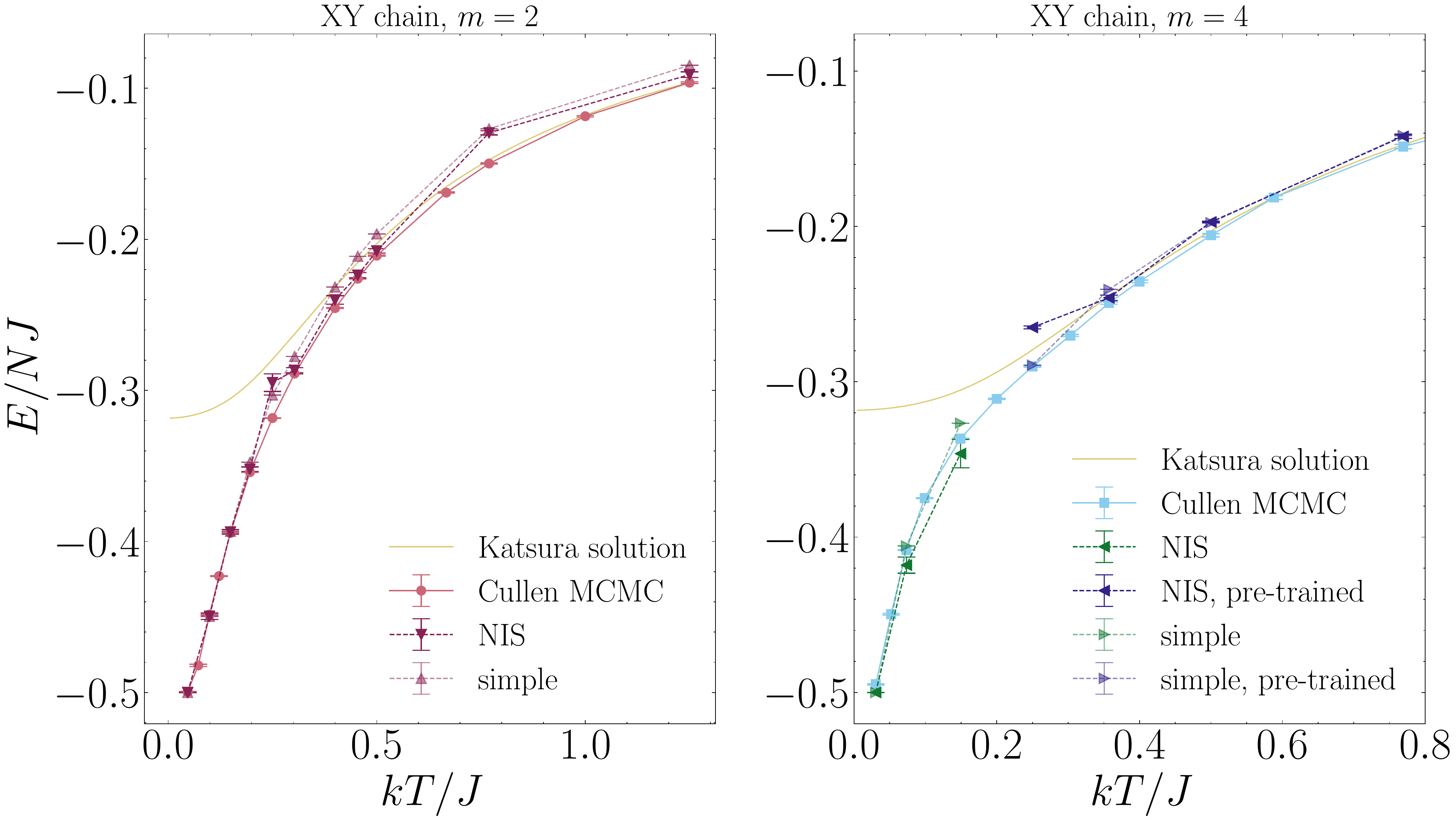}
	\caption{PixelCNN simple sample averages and NIS estimates for the internal energy $E$ of the Trotter approximation of an XY chain with $m=2$ and $m=4$ with $m=2$ (left) and $m=4$ (right). They are compared to estimates from the MCMC algorithm by Landau and Cullen \cite{mc_cullenlandau_83}, see \Cref{app:mcmc} for details. The analytical solution of Katsura \cite{xysol_katsura_62} for the quantum system is given for completeness.}
	\label{fig:XYE}
\end{figure}

\begin{figure*}[htpb]
	\centering
	\includegraphics[width=\textwidth]{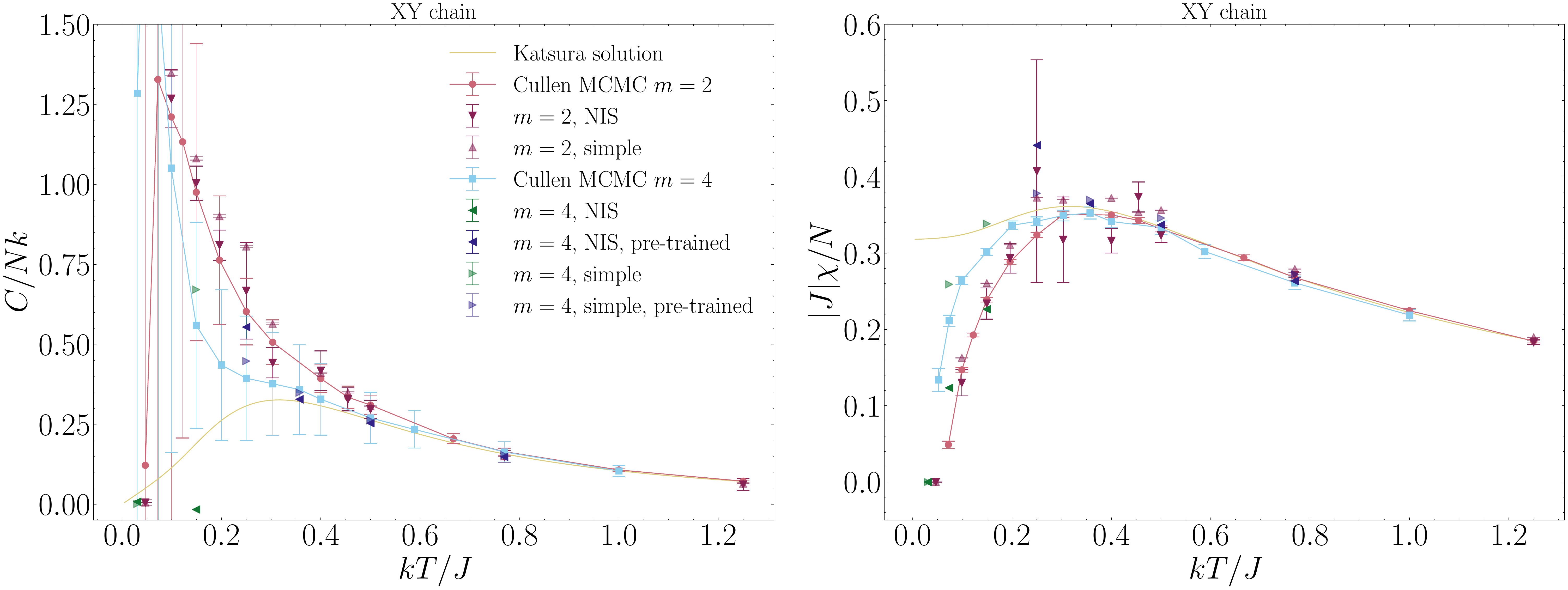}
	\caption{PixelCNN simple sample averages and NIS estimates for the specific heat $C$ (left) and susceptibility (right) of the Trotter approximation of an XY chain with $m=2$ and $m=4$. They are compared to estimates from the MCMC algorithm by Landau and Cullen \cite{mc_cullenlandau_83}, see \Cref{app:mcmc} for details. The analytical solution of Katsura \cite{xysol_katsura_62} for the quantum system is given for completeness.}
	\label{fig:XYCChi}
\end{figure*}

In \Cref{fig:XYE,fig:XYCChi}, the GNS estimates for internal energy $E$, specific heat $C$ and susceptibility $\chi$ of an XY chain with $J_x=J_y+\epsilon$ are compared to those of the MCMC algorithm described in \Cref{app:mcmc}. The analytical solution by Katsura \cite{xysol_katsura_62} is given for completeness. For this system, the difference between the ground-state of the quantum system and the classical approximation is large at low temperatures, where the quantum effects are poorly reproduced by the classical approximation with $m=2$ and $m=4$, respectively. This equally affects the GNS as well as the MCMC results.

The NIS estimates for the internal energy are in reasonable agreement with the MCMC estimates within errorbars at most temperatures for $m=2$, only at $\ktj=0.25$ the NIS estimate differs considerably and is not compatible with the MCMC result within errorbars. Here, the errors for the MCMC results were estimated according to \cite{wolff_error}. The simple sample averages for the internal energy $E$ are quite accurate at low temperatures, but deviate with increasing temperature. This reflects the difficulty of estimating $E$ at higher temperatures, which can be understood as follows: \Cref{fig:XYZ_F} shows the contributions of each plaquette type from \Cref{fig:plaquettes} to the energy function $F$ at different temperatures. At low temperatures, the contributions of plaquettes $(3)-(6)$ do not differ much. With increasing temperature, the gap between the contributions of plaquettes $(3),(4)$ and plaquettes $(5),(6)$ widens. Plaquettes $(7)$ and $(8)$ are not relevant for $J_x=J_y+\epsilon$ as they have a vanishing probability of occurring. Configurations with a high number of $(5),(6)$ plaquettes make a large contribution to the estimate while having a very low probability as can be read off from \Cref{eq:E56}. Learning to sample these rare configurations is a difficult task. This could explain the bad simple estimate at $kT=0.77$. At this temperature, also the NIS estimate does not agree with the MCMC estimate.

The specific heat $C$ of the $m=2$ classical system has a peak around $\ktj=0.07$. The NIS estimates on the descending part of the peak have a large variance but are reasonable. A comparison to the MCMC estimates is only partly possible due to their large statistical errors, which are too large to fit into the plot at temperatures below $\ktj=0.15$ and only become reasonably small at $\ktj=0.4$. At higher temperatures, all NIS estimates are compatible with the MCMC estimates and show less variance. The simple estimates for $C$ follow the MCMC estimates only qualitatively around the peak but are more accurate towards higher temperatures. The NIS estimate at $\ktj=0.05$ is poor because the model samples only configurations in the ground-state which consists of $(3)$ and $(4)$ plaquettes. This can be seen in \Cref{fig:XY_PD}, which shows the average number of plaquettes of each type across all samples.

Similar to the specific heat, the NIS estimates for the susceptibility $\chi$ have a large variance for temperatures $\ktj=0.25$ and $\ktj=0.3$ for $m=2$. At temperatures $\ktj=0.4$ and $\ktj=0.45$, the NIS estimates with error bounds do not agree with the MCMC estimates. This changes at higher temperatures, where NIS estimates are compatible with the MCMC estimates. Again, the simple sample estimates only follow the MCMC estimates qualitatively, reiterating the importance of using NIS estimates instead of simple means \cite{nicoli2019comment, kim_savan}. 

It is instructive to evaluate the effective sample sizes for the point estimate $N_{\text{ESS}}$ in \Cref{eq:ESS} and its variance $N_{\text{ESS},\sigma}$  as defined in \Cref{eq:ESS_err} along the lines of \cite{mcbook}, which are shown in \Cref{fig:ESS}.
$N_{\text{ESS}}$ is particularly small for $\ktj=0.25$ and $\ktj=0.3$ for $m=2$. This could explain why the NIS estimates for $E$ and $\chi$ at these temperatures are poor. Most notably, $N_{\text{ESS},\sigma}$ is only one at $\ktj=0.3$. The lower limit for the effective sampling sizes above which the estimate is still trustworthy is not clear. Although $N_{\text{ESS}}$ and $N_{\text{ESS}, \sigma}$ at $\ktj=0.4$\;and $\ktj=0.45$\;are not as low as at other temperatures for $m=2$, the NIS estimates of $\chi$ are bad at these temperatures. However, the effectiveness of importance sampling also depends on the function that is sampled which is not taken into account by $N_{\text{ESS}}$ and $N_{\text{ESS},\sigma}$.

For $m=4$, the estimates of a PixelCNN model trained as outlined in \Cref{app:exp_det} are reasonable up to only $\ktj=0.15$. At higher temperatures, the model collapses into either sampling only ground-state configurations or exclusively configurations of only $(1)$ or $(2)$ plaquettes. This can be seen in \Cref{fig:XY_PD}. Fortunately, the PixelCNN could be prevented from collapsing by using a pretraining schedule. For the model at $\ktj=0.25$, training of the model at $\ktj=0.15$ was continued for \SI{2000}{} epochs. The resulting model was then used as a pretrained model for the one at $\ktj=0.36$. This procedure was iterated up to $\ktj=0.77$.

\subsection{Isotropic (XXX) chain}

\begin{figure}[htpb]
	\centering
	\includegraphics[width=\columnwidth]{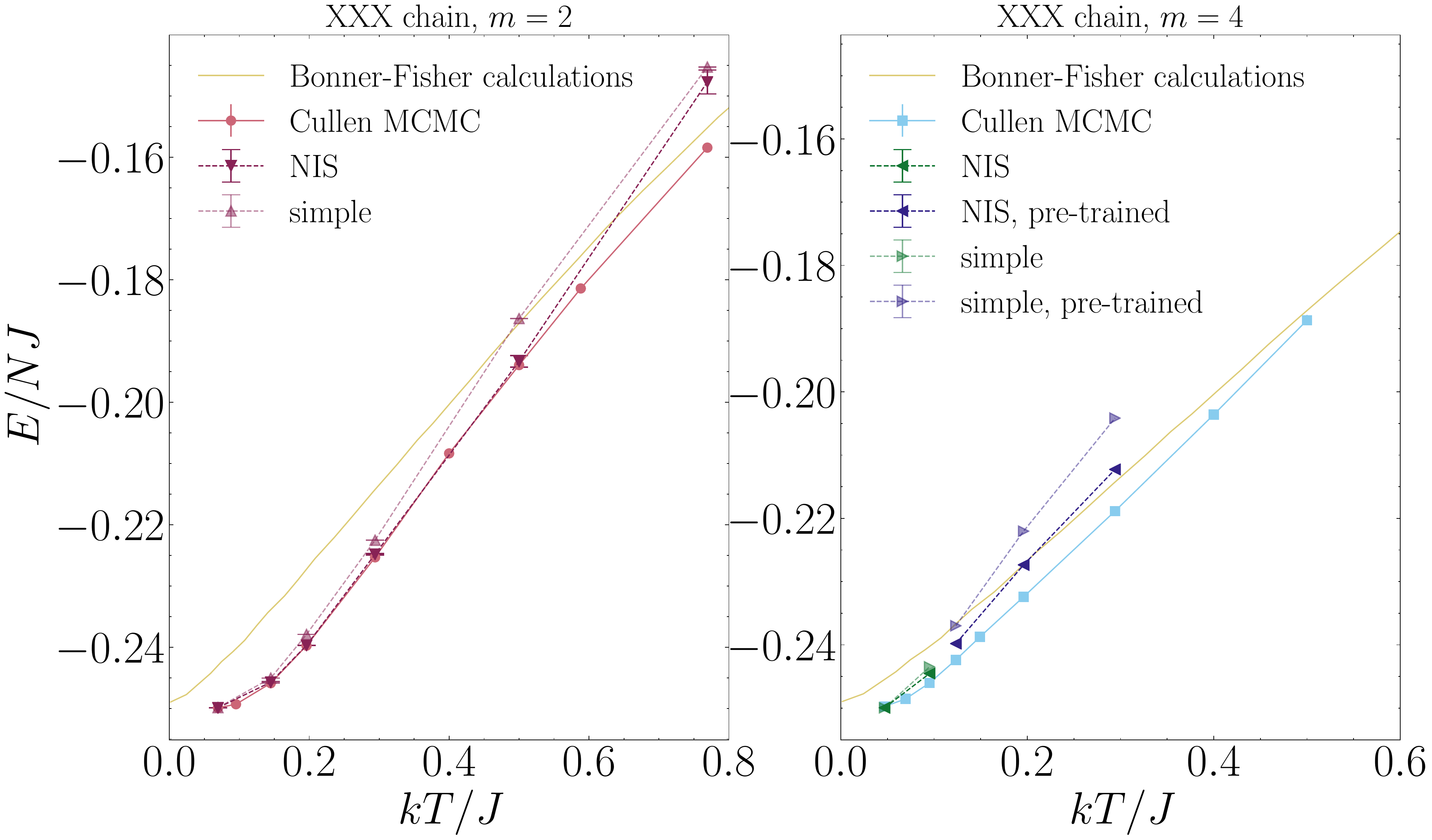}
	\caption{PixelCNN simple sample averages and NIS estimates for the internal energy $E$ of the Trotter approximation of an almost isotropic Heisenberg chain with $m=2$ (left) and $m=4$ (right). They are compared to estimates from the MCMC algorithm by Landau and Cullen \cite{mc_cullenlandau_83}. The results of the finite-size calculations by Bonner and Fisher \cite{calc_bonnerfisher_64} for the quantum system are given for completeness.}
	\label{fig:XYZE}
\end{figure}

\begin{figure*}[htpb]
	\centering
	\includegraphics[width=\textwidth]{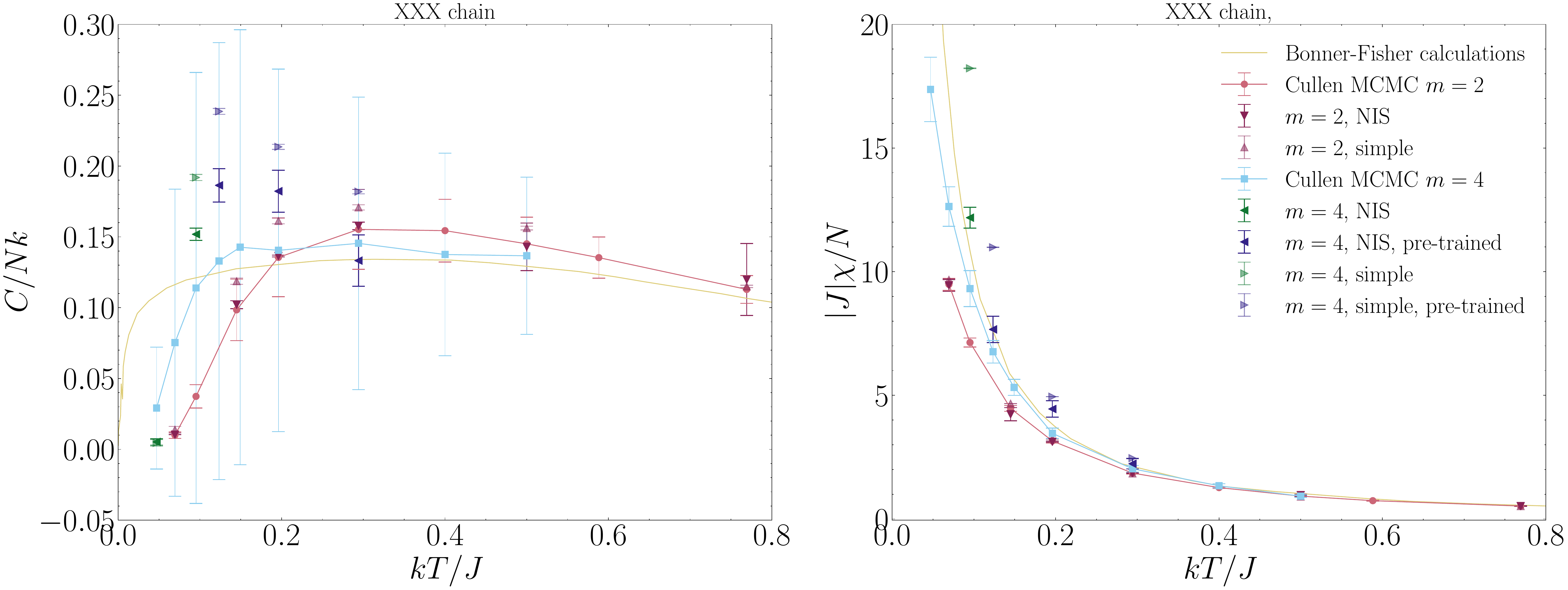}
	\caption{PixelCNN simple sample averages and NIS estimates for the specific heat $C$ (left) and susceptibility $\chi$ (right) of the Trotter approximation of an almost isotropic Heisenberg chain with $m=2$ and $m=4$. They are compared to estimates from the MCMC algorithm by Landau and Cullen  \cite{mc_cullenlandau_83}. The results of the finite-size calculations by Bonner and Fisher \cite{calc_bonnerfisher_64} for the quantum system are given for completeness.}
	\label{fig:XYZCChi}
\end{figure*}

\Cref{fig:XYZE,fig:XYZCChi} show PixelCNN estimates for the internal energy $E$, specific heat $C$ and susceptibility $\chi$ of an almost isotropic chain with $J_x=J_y+\epsilon$ approximated by a classical system with $m=2$ and $m=4$. They are compared to those of the MCMC algorithm described in \Cref{app:mcmc}. The finite-size calculations for the quantum system by Bonner and Fisher \cite{calc_bonnerfisher_64} are given for reasons of completeness. The ground-state energy of the quantum system and the classical approximation are equal. This is why for low temperatures, the MCMC results are quite close to the Bonner-Fisher-curve, considering the small extension in Trotter direction of the classical approximation.

For $m=2$ the PixelCNN model performs well in estimating the internal energy $E$, specific heat $C$ and susceptibility $\chi$ up to a certain temperature. The NIS estimates agree with the MCMC estimates within error bounds for all observables and temperatures except for $E$ at $\ktj=0.76$.
The simple sample mean estimates only follow the MCMC estimates for $E$ and $C$ qualitatively. For $E$, the deviation between these estimates increases with temperature. 

Interestingly, for $\chi$, the simple estimates are quite close to the MCMC estimates. For this observable, only plaquettes $(1),(2)$\;contribute to the sum.  Apparently, the models are good at sampling the correct ratio of these plaquettes. 
The bad estimate at $\ktj=0.77$ for only $m=2$ already reflects the difficulty of estimating $E$ at higher temperatures described in \Cref{sec:XYchain}.

The question remains why the NIS estimator at $\ktj=0.77$ for $m=2$ is not much better than the simple average. Again, it is informative to look at the effective sample sizes for the point estimate $N_{\text{ESS}}$ in \Cref{eq:ESS} and its variance in \Cref{eq:ESS_err} shown in \Cref{fig:ESS}. For $m=2$, $N_{\text{ESS}}$ is considerably lower than the sample size of \SI{4e6}\;for all temperatures. For $\ktj=0.14$ it is very  small. Most notably, $N_{\text{ESS}, \sigma}$ is close to only one at $\ktj=0.77$, which means this variance estimate is not trustworthy. Thus, the distribution $q_{\theta}$ learned by the model is not close enough to the Boltzmann distribution $p$ to estimate $E$ at $\ktj=0.77$ with a reasonable amount of samples. Although the effective sampling sizes at $\ktj=0.14$\;are small, the NIS estimates at this temperature agree with the MCMC estimates. As mentioned above, $N_{\text{ESS}}$ and $N_{\text{ESS}, \sigma}$ do not depend on the form of the function that is sampled although the effectiveness of importance sampling depends on it. For the observables considered here, the learned distribution $q_{\theta}$ at $\ktj=0.14$\;seems to be sufficient for importance sampling.

Similar to the XY chain, the PixelCNN model collapses when trained from scratch at temperatures above $\ktj=0.1$. A pretraining schedule like for the XY chain is employed for the models at higher temperatures. However, the estimates still follow the MCMC results only qualitatively.

\section{Conclusions and Outlook}
\label{sec:conclusions}
This work explores generative neural samplers to evaluate thermodynamic observables in quantum mechanical statistical model as a complementary approach to Monte Carlo methods. An autoregressive model is applied to a quantum Heisenberg chain approximated by a two-dimensional classical spin system with complicated four-spin interactions. The resulting estimates for energy, specific heat and susceptibility for the isotropic chain and the XY chain are in good agreement with Monte Carlo results. Neural Importance Sampling was used to correct for the unavoidable sampling bias of the generative models. It is key to achieve not only qualitative but also quantitative agreement with MCMC results. A major difficulty at higher temperatures is that configurations with vanishing probability give high contributions to the non-trivial thermal average of the internal energy. This becomes more severe with larger Trotter length $m$. Currently, systems with Trotter length $m=4$ are the limit of this approach, where even pretraining on lower temperatures had to be used to prevent the models from collapsing. Using a carefully designed pretraining protocol, the PixelCNN might succeed on systems with larger $m$.

From the machine learning perpective, application of generative models to real-world physical systems represents a major challenge not only in high-energy physics \cite{lft_u1,lft_SUN} but also in solid-state physics, where this work might serve as a starting point for future investigations.
A particular advantage over generative modeling in other domains is the availability of unambiguous ground truth data to evaluate the quality of the sampling algorithm. 

The central challenge in all of the physical application domains mentioned above, is the correct incorporation of symmetries. Previously, autoregressive models have been applied to the Ising model which only has a discrete $Z(2)$ symmetry. The Heisenberg chain shows $U(1)$ and $SU(2)$ symmetry for the case of the XY chain and the ferromagnetic isotropic chain, respectively. 
In the classical approximation, the difficulty of learning a continuous symmetry reflects in the complicated four spin interactions, which only allow configurations with eight types of four-spin plaquettes. This was in our case accounted for by a modified sampling procedure. Likely, learning to incorporate periodic boundary conditions is difficult for architectures that process information in a sequential manner such as the PixelCNN.
Transformer-based architectures such as iGPT \cite{iGPT} represent a promising direction for future research because they allow the network to directly connect information from two arbitrary sites in the lattice.

\section*{Acknowledgements}
This work was supported by the Bundesministerium f\"ur Bildung und Forschung through the BIFOLD - Berlin Institute for the Foundations of Learning and Data (ref. 01IS18025A and ref. 01IS18037A).

\appendix

\section{Katsura's solution for the XY chain } 
\label{ch:katsura_sol}
The internal energy $E$, the specific heat $C$ and the susceptibility $\chi$ derived from Katsura's solution \cite{xysol_katsura_62} for the free energy $F$ of the XY chain in \Cref{eq:xyF} are given explicitly by
\begin{equation} \label{eq:xyE}
	E = -\frac{NkT}{\pi} \int_0^{\pi}  \sqrt{h(\omega)} \tanh \left(\sqrt{h(\omega)} \right) \,d\omega\,,
\end{equation}
and
\begin{equation} \label{eq:xyC}
	C = \frac{N}{\pi} \int_0^{\pi} \frac{h(\omega)}{\cosh \left(  \sqrt{h(\omega)}  \right)^2 } \,d\omega\,,
\end{equation}
and
 \begin{align} \label{eq:xyChi}
	\chi = -\frac{N}{4 \pi kT} \int_0^{\pi} &\frac{\tanh \left( \sqrt{h(\omega)} \right)}{\sqrt{h(\omega)}} 
	+ \frac{ \left[ \left(K_x + K_y \right)  \cos\omega \right]^2 }{h(\omega) \cosh \left( \sqrt{h(\omega)} \right)^2 } \notag\\
	&- \frac{\left[ \left(K_x + K_y \right) \cos\omega \right]^2  \tanh\left(\sqrt{h(\omega)} \right)^2 }{h(\omega)^{3/2}} \,d\omega \,,
\end{align}
with $h(\omega) = K_x^2 +K_y^2 +2K_x K_y \cos(2\omega)$ and $K_x=\frac{J_x}{4kT}$, $K_y=\frac{J_y}{4kT}$.

\section{Derivation of the Suzuki-Trotter transformation} \label{sec:st_derivation}
In this section, we recapitulate the derivation of the Suzuki-Trotter transformation for the Heisenberg chain. According to the Suzuki-Trotter transformation \cite{st_formula_trotter,st_formula_suzuki} a set of bounded operators $\left \{ A_j \right \}$ fulfills the relation
\begin{equation} 
	\exp \left( \sum_{j=1}^p A_j \right) = \lim \limits_{m \to \infty} \left( e^{A_1 / m} e^{A_2 / m} \dots e^{A_p / m} \right)^m\,. \label{eq:st_trafo}
\end{equation}
\Cref{eq:st_trafo} can be applied to a quantum statistical mechanical system after its Hamiltonian $\ham$ is decomposed into a sum of terms $\ham = \ham_1+\ham_2+ \dots + \ham_p$. Its partition function,
\begin{equation}
	Z = \Tr  e^{-\beta \ham} \,,
\end{equation}
can be approximated by 
\begin{equation} 
	Z^{(m)} = \Tr ( e^{-\beta \ham_1/m} e^{-\beta \ham_2/m} \dots e^{-\beta \ham_p/m} )^m\,. \label{eq:Z_decomp}
\end{equation} 
The convergence of the approximation to $Z$ as $m \to \infty$ is studied in \cite{trotter_conv}.
There are several ways to split up the Hamiltonian of the quantum Heisenberg chain \Cref{eq:qhc_hamiltonian} in order to apply \Cref{eq:Z_decomp}. Here, the so-called checkerboard decomposition is used \cite{checkerboard_barma_78}
\begin{equation} \label{eq:checkerboard}
	\ham = \ham_0 + V_A + V_B	\,,
\end{equation}
where
\begin{align}
	\ham_0 &= - \sum^N_{i=1}J_z S_i^z S_{i+1}^z\,, \label{eq:H0} \\
	%V_A &= \sum^{N-1}_{i=1} V_i \; \text{with odd}\, i\, \text{values} \\
	V_A &= \sum_{i\;\text{odd}} V_i, \;\; V_B = \sum_{i\;\text{even}} V_i\,,  \label{eq:VaVb}\\
	%V_B &= \sum^{N}_{i=2} V_i \; \text{with even}\, i \,\text{values} \
	V_i &= - \left( J_x S_i^x S_{i+1}^x + J_y S_i^y S_{i+1}^y \right)\,.
\end{align}
Applying \Cref{eq:Z_decomp} one obtains
\begin{equation}
	Z^{(m)} = \Tr \left( e^{-\beta \ham_0/2m} e^{-\beta V_A/m} e^{-\beta H_0/2m} e^{-\beta V_B/m} \right) ^m\,.
\end{equation}
Now, $2m$ complete sets of eigenstates of $\ham_0$ are inserted, so that there is one complete set $\{\alpha\}$ between each exponential, which yields
\begin{multline} \label{eq:trace_operation}
	Z^{(m)} =\!\!\! \sum_{\alpha_1 \alpha_2 \dots \alpha_{2m}}\!\!\! \exp \left(-\frac{\beta}{2m} \sum_{r=1}^{2m}\ham_0^r \right)  \bra{\alpha_1} e^{-\beta V_A/m} \ket{\alpha_2} \\ \bra{\alpha_2} e^{-\beta V_B/m} \ket{\alpha_3} \dots \bra{\alpha_{2m}} e^{-\beta V_B/m} \ket{\alpha_1}\,.
\end{multline}
$r$ in the superscript of $\ham_0$ is just a dummy integer to keep track of the terms, meaning $\ham_0\ket{\alpha_r}=\ham_0^r\ket{\alpha_r}$. Each $\alpha_r$ runs over $2^N$ states. Because $\ket{\alpha_r}$ are eigenstates of $\ham_0$,
\begin{equation}
S_i^z \ket{\alpha_r} = S_{i,r} \ket{\alpha_r}, \;\; S_{i,r}=\pm\frac{1}{2} \,.
\end{equation}
Inserting the operators $\ham_0$ from \Cref{eq:H0} and $V_A$ and $V_B$ from \Cref{eq:VaVb} one obtains \cite{mc_cullenlandau_83}
\begin{multline} \label{eq:4spin_ising_app}
	Z^{(m)} = \sum_{\alpha_1 \alpha_2 \dots \alpha_{2m}} \exp \Bigl( 2K_0 \sum^N_{i=1}\sum_{r=1}^{2m}S_{i,r}S_{i+1,r}  \\  - \beta \sum_{\hat i}\sum_{\hat r} h(i,r) \Bigr)\,.
\end{multline}
The caret denotes that the sum is running over terms with $i$ and $r$ both odd or both even. It is $K_0 = \frac{\beta}{4m}J_z$ and 
\begin{align} \label{eq:4spin_matrixel}
	h&(i,r) =\notag\\
	& -\frac{1}{\beta} \ln \bra{S_{i,r}, S_{i+1,r}} \exp\left( -\beta V_i/m \right) \ket{S_{i,r+1}, S_{i+1,r+1}}\,.
\end{align}
These matrix elements can be found using the identity 
\begin{align}
	\exp \left( -\beta V_i/m \right) &= \left( \frac{1}{2} + S_i^zS_{i+1}^z \right) \cosh K_{-} \nonumber \\  &+ \left( \frac{1}{2} - 2S_i^zS_{i+1}^z \right) \cosh K_{+} \nonumber\\
		&+ \left( S_i^+S_{i+1}^- + S_{i+1}^+S_i^- \right) \sinh K_{+} \nonumber \\
		 & + \left( S_i^+S_{i+1}^+ + S_{i+1}^-S_i^- \right) \sinh K_{-} \,,
\end{align}
where $K_{\pm} = \frac{\beta}{4m} \left( J_x \pm J_y \right)$.

\section{Derivation of thermal averages} \label{sec:thermal_av_derivation}
The thermal properties of the trotterized Heisenberg chain, the respective derivatives of the free energy are taken. To obtain them in a convenient form, the partition function is rewritten as
\begin{align}
	Z^{(m)} = \sum_j \exp(-\beta E_j^{(m)})\,,
\end{align}
where $E_j^{(m)}$ is the energy of the $j$th state obtained by summing the energy contributions of its 4-spin plaquettes from \Cref{eq:E12,eq:E34,eq:E56,eq:E78}. The thermal average of the \textit{internal energy} is
\begin{align}
	E^{(m)} &= -\frac{\partial}{\partial \beta} \ln Z^{(m)} \notag\\
		   &= \frac{1}{Z^{(m)}} \sum_j \left[ \frac{\partial}{\partial \beta} (\beta E_j^{(m)})  \right] \exp \left( -\beta E_j^{(m)} \right) \notag\\
		   &= \frac{1}{Z^{(m)}} \sum_j F_j^{(m)} \exp \left( -\beta E_j^{(m)} \right)\,.
\end{align}
$F_j^{(m)}$ is the non-trivial \textit{energy function} in the $j$th state whose ensemble average is the thermal average, i.e.\
\begin{equation}
	E^{(m)} = \langle F^{(m)} \rangle\,.
\end{equation}
From here, the \textit{specific heat} is found by using the expression
\begin{equation}
	C^{(m)} = - \frac{1}{T^2} \frac{\partial E^{(m)}}{\partial \beta}\,,
\end{equation}
which gives
\begin{equation}
	C^{(m)} = \frac{1}{T^2} \left( \langle F^{(m)}{}^2 - G^{(m)} \rangle - \langle F^{(m)} \rangle^2 \right)\,,
\end{equation}
with
\begin{equation}
	G_j^{(m)} = \frac{\partial^2}{\partial \beta^2} \left( \beta E_j^{(m)} \right)\,.
\end{equation}
Finally, to use the expression for the \textit{parallel susceptibility}
\begin{equation} \label{eq:chi_exp}
	\chi^{(m)} = - \frac{\partial^2}{\partial H^2} \left(-\frac{1}{\beta}\ln Z^{(m)} \right) \Biggr|_{H=0}\,,
\end{equation}
the Suzuki-Trotter transformation has to be applied to the partition function of the model with an external field $H$ applied in the $z$ direction. Thus, the term $H\sum_iS_i^z$ is added to the Hamiltonian in \Cref{eq:qhc_hamiltonian}. The transformation of the resulting partition function is analogous to the one described in the previous section and the result is
\begin{equation}
	Z_H^{(m)} = \sum_j \exp \left( -\beta E_j^{(m)} + \frac{\beta H}{2m} \sum_{i=1}^N \sum_{r=1}^{2m} S_{i,r}\right)\,.
\end{equation}
Applying \Cref{eq:chi_exp} the susceptibility is obtained as 
\begin{equation}
	\chi^{(m)} = \beta \left( \langle M^{(m)^2} \rangle - \langle M^{(m)} \rangle^2 \right)\,,
\end{equation}
where 
\begin{equation}
	M_j^{(m)} = \frac{1}{2m} \left( \sum_i\sum_r S_{i,r} \right)\,.
\end{equation}

The values for the quantities $E^{(m)}$, $F^{(m)}$, $G^{(m)}$, $M^{(m)}$ for a single 4-spin plaquette are given below for each allowed plaquette of the general anisotropic Heisenberg chain approximated by a classical Ising-like model with $m$:

\begin{align} 
	E^{(m)}(1,2) &= - \frac{1}{\beta} \left( K_0 + \ln \cosh K_- \right)\,, \label{eq:E12_a} \\
	F^{(m)}(1,2) &= - \frac{1}{\beta} \left( K_0 + K_- \tanh K_- \right)\,,\label{eq:F12} \\
	G^{(m)}(1,2) &= - \left( \frac{1}{\beta} K_- \sech K_- \right)^2\,, \label{eq:G12} \\
	M^{(m)}(1,2) &= \pm \frac{1}{2m}\,, \label{eq:M12}
\end{align}
\begin{align} 
	E^{(m)}(3,4) &=   \frac{1}{\beta} \left( K_0 - \ln \cosh K_+ \right)\,, \label{eq:E34_a}\\
	F^{(m)}(3,4) &=   \frac{1}{\beta} \left( K_0 - K_+ \tanh K_+ \right)\,, \label{eq:F34}\\
	G^{(m)}(3,4) &=   - \left( \frac{1}{\beta} K_+ \sech K_+ \right)^2\,, 
	\label{eq:G34} \\
	M^{(m)}(3,4) &= 0\,, \label{eq:M34}
\end{align}
\begin{align} 
	E^{(m)}(5,6) &=   \frac{1}{\beta} \left( K_0 - \ln \sinh K_+ \right)\,, \label{eq:E56_a}\\
	F^{(m)}(5,6) &=   \frac{1}{\beta} \left( K_0 - K_+ \coth K_+ \right)\,, \label{eq:F56}\\
	G^{(m)}(5,6) &=   \left( \frac{1}{\beta} K_+ \csch K_+ \right)^2\,, 
	\label{eq:G56}\\
	M^{(m)}(5,6) &= 0\,, \label{eq:M56}
\end{align}
\begin{align} 
	E^{(m)}(7,8) &= - \frac{1}{\beta} \left( K_0 + \ln \sinh K_- \right)\,,\label{eq:E78_a}  \\	
	F^{(m)}(7,8) &= - \frac{1}{\beta} \left( K_0 + K_- \coth K_- \right)\,,\label{eq:F78} \\	
	G^{(m)}(7,8) &=   \left( \frac{1}{\beta} K_- \csch K_- \right)^2\,, \label{eq:G78}\\
	M^{(m)}(7,8) &= 0\,. \label{eq:M78}
\end{align}
\begin{figure}[htpb]
	\centering
	\includegraphics[width=\columnwidth]{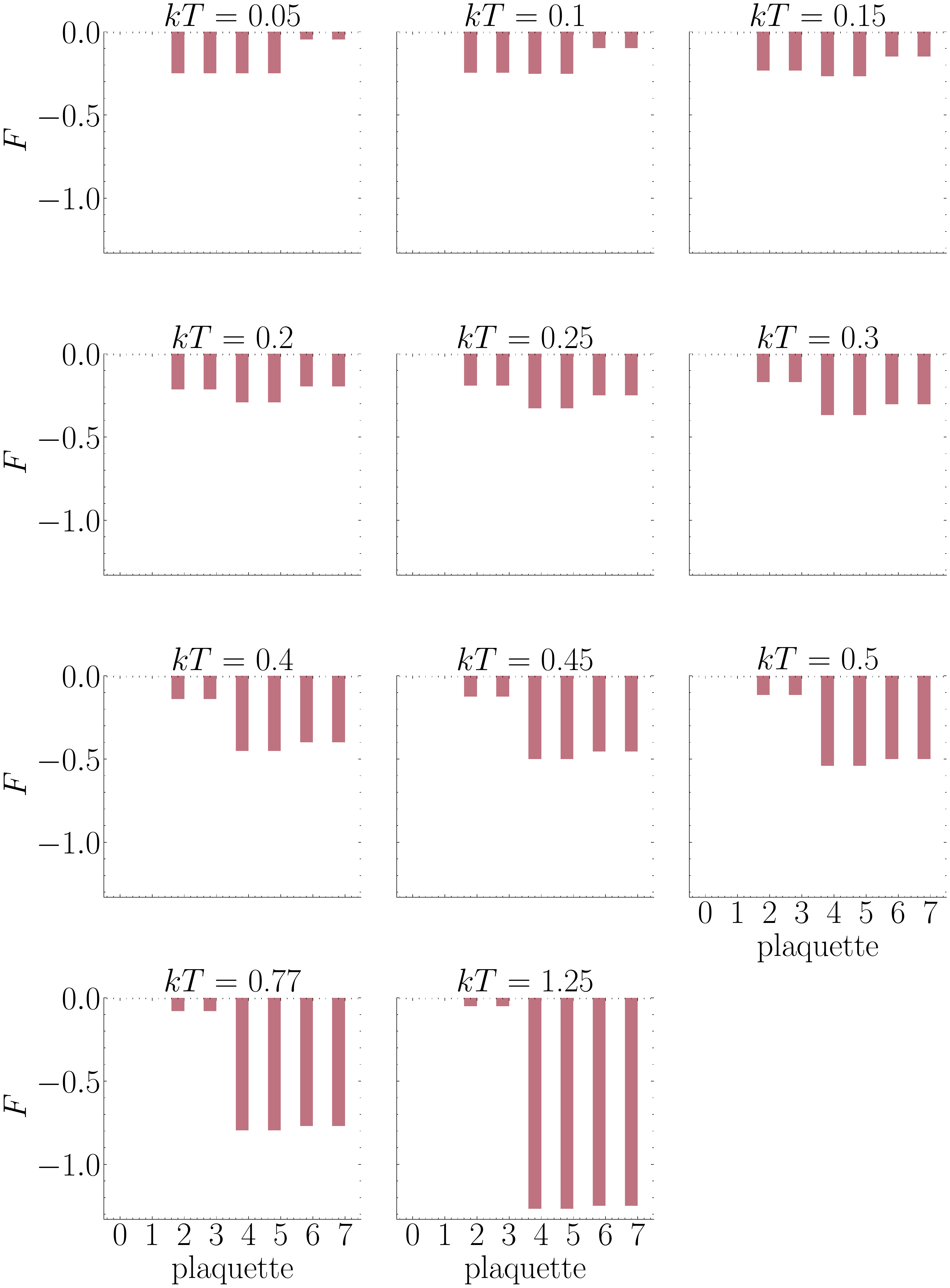}
	\caption{Contributions $F^{(2)}(i)$ to the energy function of each plaquette type $(i)$. Example for a XY Heisenberg chain approximated by a classical Ising-like model with $m=2$}
	\label{fig:XYZ_F}
\end{figure}

\section{Neural Importance Sampling (NIS)}
As a matter of fact, the sampling distribution $q_{\theta}$ will never match the Boltzmann distribution $p$ perfectly, i.e.\ there is an inevitable sampling error. Thus, evaluating expectation values of observables via the simple sample mean,
\begin{equation}
	\langle \obs (s) \rangle_p \approx \frac{1}{N} \sum_{i} \obs(s_i) \label{eq:simple_mean}
\end{equation}
leads to a systematic error and when reporting simple sample means, one has essentially no control over the quality of the approximation.
To remedy this problem, \cite{nicoli2019comment,kim_savan} proposed a sampling method that guarantees asymptotical unbiasedness named Neural Importance Sampling (NIS). The NIS Estimator is
\begin{equation} 
	  \widehat{\obs}_N =  \sum_i w_i \obs(s_i) \approx  \langle \obs (s) \rangle_p\,, \label{eq:NISestimator}
\end{equation}
with the normalized importance weights $w_i = \frac{\hat w_i}{\sum_i \hat w_i}$ where $\hat w_i = \frac{e^{-\beta \ham(s_i)}}{q_\theta(s_i)}$ are the unnormalized weights and $s_i \sim q$ are independent identically distributed samples. For this sampling method, one has to ensure that the support of the sampling distribution $q_\theta$ contains the support of the target distribution $p$. To achieve this, the original model output $q' \in [0,1]$ is rescaled \cite{nicoli2019comment,kim_savan} according to
\begin{equation}
	q = (q' -\frac{1}{2})(1-2\epsilon)+ \frac{1}{2}\,,
\end{equation}
such that the probability $q$ lies in $[\epsilon,1-\epsilon]$. 

Neural importance sampling comes with the advantage of well founded estimates for the variance of a point estimate $\widehat{\obs}_N$. The delta method is applied to \Cref{eq:NISestimator} to get an approximate variance for the NIS estimate \cite{mcbook}: 

\begin{equation}
	\text{Var} \left( \widehat{\obs}_N \right) = \frac{1}{N} \frac{\E_q \left[ \left( \obs(s) \frac{p(s)}{q_\theta(s)} - \E_p[\obs] \frac{p(s)}{q_\theta(s)} \right)^2 \right]}{\E_q \left[ \frac{p(s)}{q_\theta(s)} \right]^2}\,,
\end{equation}

which can be estimated by

\begin{align}
    \widehat{\text{Var}} \left( \widehat{\obs}_N \right) &= 
    \frac{ \frac{1}{N} \sum_{i=1}^N \widehat w(s_i)^2 \left( \obs(s_i) - \widehat{\obs}_N \right)^2}
    { \frac{1}{N} \left( \sum_{i=1}^N \widehat w(s_i) \right)^2} \\
    &= \sum_i^N w_i^2 \left( \obs (s_i) - \widehat{\obs}_N \right)^2\,.     
\end{align}

However, it is not guaranteed that a poor point estimate $\widehat{\obs}_N$ will have a large variance $\widehat{\text{Var}}\left( \widehat{\obs}_N \right)$. The variance estimate is based on the same unequal weights as the point estimate. If the weights are too skewed, not only is the point estimate bad, but also the variance estimate is not reliable. The effective sampling size \cite{mcbook}

\begin{equation} \label{eq:ESS}
	N_{\text{eff}} = \frac{N}{\sum_{i=1}^N w_i^2}
\end{equation}
is a diagnostic for problematic weights concerning the point estimate.
If $N_{\text{eff}}\ll N$, the weights are considerably imbalanced and the estimate is similar to an average of only $N_{\text{eff}}$ samples. 
An analogous measure for the variance estimate $\widehat{\text{Var}}$ which depends on weights $w_i^2$ is given by

\begin{equation} \label{eq:ESS_err}
	N_{\text{eff},\sigma} = \frac{\left( \sum_{i=1}^N w_i^2 \right)^2}{\sum_{i=1}^N w_i^4}\,.
\end{equation}

\begin{figure}[htpb]
	\centering
	\includegraphics[width=\columnwidth]{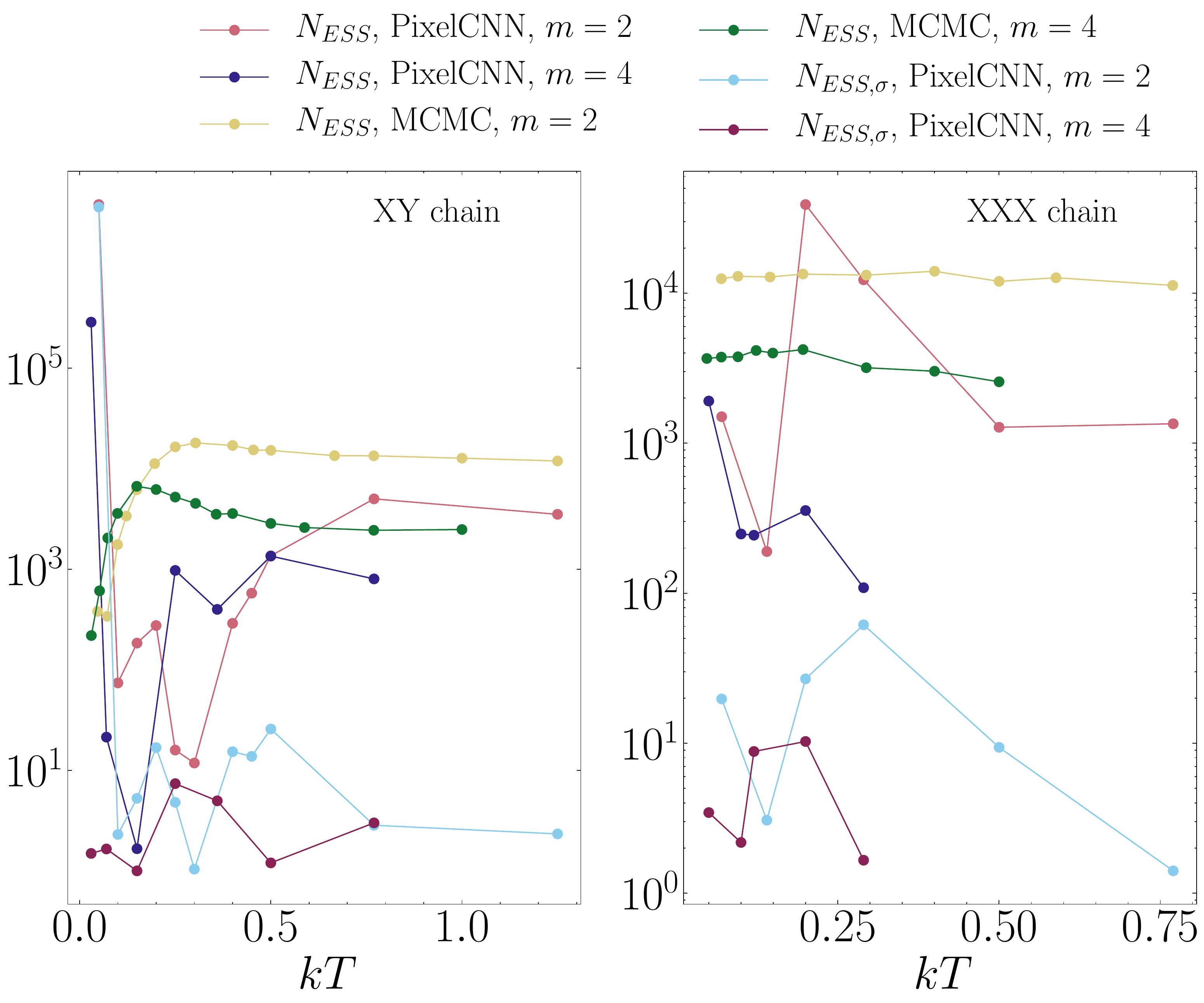}
	\caption{Effective sample sizes for the NIS point estimate $N_{\text{ESS}}$ in \Cref{eq:ESS} and its variance $N_{\text{ESS}, \sigma}$ in \Cref{eq:ESS_err}. The effective sampling size of the MCMC estimates defined as $N_{\text{ESS}}=\frac{N}{2\tau_{int,\obs}}$ is plotted for the internal energy $E$. $\tau_{int}$ is the integrated autocorrelation time of the operator $\obs$.\cite{wolff_error} Left: XY chain. Right: almost isotropic chain}
	\label{fig:ESS}
\end{figure}

If $N_{\text{eff},\sigma}$ is small, the variance estimate cannot be trusted. However, $N_{\text{eff}}$ and $N_{\text{eff},\sigma}$ are imperfect diagnostics, because it highly depends on the application above which values of them an estimate can be trusted. \Cref{fig:ESS} shows $N_{\text{ESS}}$ and  $N_{\text{ESS},\sigma}$  for the estimates discussed in \Cref{sec:results}.

\section{MCMC algorithm}
\label{app:mcmc}
\begin{figure}[htpb]
	\centering
	\includegraphics[width=\columnwidth]{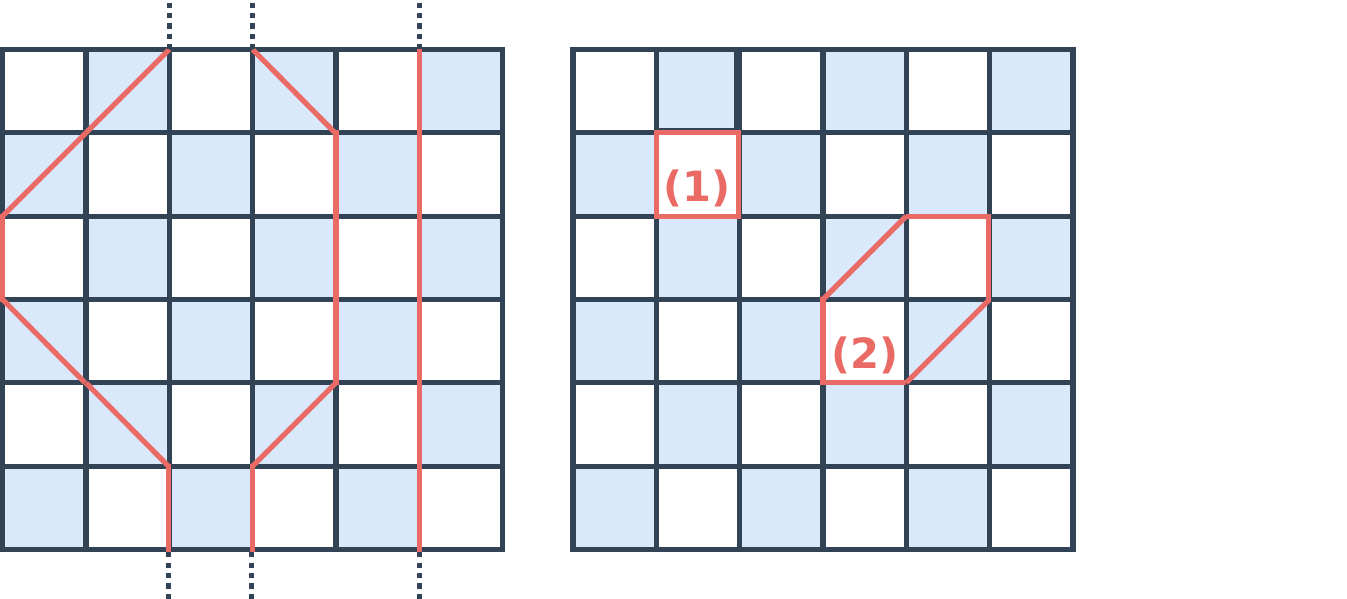}
	\caption{Left: zigzag flipping patterns on the effective lattice for $m=3$ - the strings close due to the periodic boundary conditions. Right: Local flipping patterns}
	\label{fig:flipping_patterns}
\end{figure}

\cite{mc_cullenlandau_83} carried out Monte Carlo simulations of quantum Heisenberg chains mapped to two-dimensional Ising systems with 4-spin interactions like in \Cref{sec:suzuki_trotter}. The classical Metropolis can be employed to simulate the Ising like system only with an adapted spin flipping method. The spin flipping pattern is the major difficulty, as the mapped Ising model only allows configurations with the 4-spin plaquettes in \Cref{fig:plaquettes}. Flipping an arbitrary pattern of spins will in general not lead to a transition from one allowed state to another. The plaquettes can be transformed into each other by flipping either two or four spins. As each spin is shared between two plaquettes on the grid, flipping two spins of a particular plaquette results in illegal configurations on the two adjacent plaquettes. In order to obtain an allowed state, a closed string of spins must be flipped, i.e. the first and the last flipped spin lie on the same plaquette. \cite{mc_cullenlandau_83} used two kinds of closed strings: zig-zag and local patterns shown in \Cref{fig:flipping_patterns}. The zigzag strings are built of only vertical and diagonal steps and start at the bottom line. Lateral steps are not allowed. The flipping algorithm is restricted to the smaller pattern (1) of the local patterns because pattern (2) and all larger patterns can be constructed by merging several patterns of type (1). These restrictions were validated in a comparison with a more complicated unrestricted flipping algorithm. In this second algorithm, vertical, lateral and diagonal steps and a random starting point were allowed. The string is built up until it intersects itself and the tail is discarded to obtain a closed string. The two algorithms gave similar numerical results.
The restricted flipping algorithm is outlined in \Cref{alg:flipping}. In \cite{mc_cullenlandau_83}, each zigzag pattern is constructed step by step randomly with increasing bias that ensures that the string is closed. In the implementation of this work, all possible zigzag patterns are constructed and a random uniform choice is made among them. All possible zigzag patterns are found in a brute-force search.
 
\begin{algorithm}
\caption{Spin flipping algorithm by \cite{mc_cullenlandau_83}}
\label{alg:flipping}
\begin{algorithmic}[1]
\State{Initialize a state with all spins up}
\Repeat
\State{Choose a random point $i'$ along the real direction}
\State{Make a uniform random choice between the local square pattern or the zigzag pattern type}
\If{Local square pattern}
\State{Choose a random point $r'$ along the Trotter direction}
\State{Flip the spins along the square pattern starting at $(i',r')$}
\ElsIf{Zigzag pattern}
\State{Construct a zigzag pattern starting at $i'$ randomly}
\State{Flip the spins along the zigzag string}
\EndIf
\Until{$N$ trials are reached for one Monte Carlo step}
\end{algorithmic}
\end{algorithm}
\section{Experimental details}
\label{app:exp_det}
The hyperparameter setup that was used to train the PixelCNN models is shown in \Cref{tab:hyperparam}. A PixelCNN architecture with residual connections and a kernel radius of \SI{3} was employed. On a Nvidia Tesla V100 GPU, training took \SI{7}{\hour} for $m=2$ and \SI{13.7}{\hour} and for $m=4$. For the estimates, \SI{4e6} samples were generated which took \SI{5}{\hour} for $m=2$ and \SI{10}{\hour} for $m=4$. The MCMC algorithm was implemented as described in \Cref{app:mcmc}. After \SI{1000}\,equilibration steps, \SI{3 e4}\;Monte Carlo steps were used for the $m=2$ trotterized Heisenberg chain and \SI{4 e4}\;Monte Carlo steps were used for $m=4$. The Markov chain was initialized in the ground states of either model, which is a state of either only $(1)$ or $(2)$ plaquettes for the almost isotropic chain, and a state of only $(3), (4)$ plaquettes for the XY chain.
\begin{table}[ht]
\centering
\begin{tabular}{ll}
	parameter & value\\
\toprule
	sampler & PixelCNN\\
depth & 6\\
 width & 64\\
  batch size & 512\\
  learning rate & 0.001\\
  steps & 10000\\
  parameters & $826369$\\
\end{tabular}
\caption{Hyperparameter setup for the PixelCNN model}
\label{tab:hyperparam}
\end{table}

\section{Plaquette distributions}
\begin{figure*}[htpb]
	\centering
	\includegraphics[width=\textwidth]{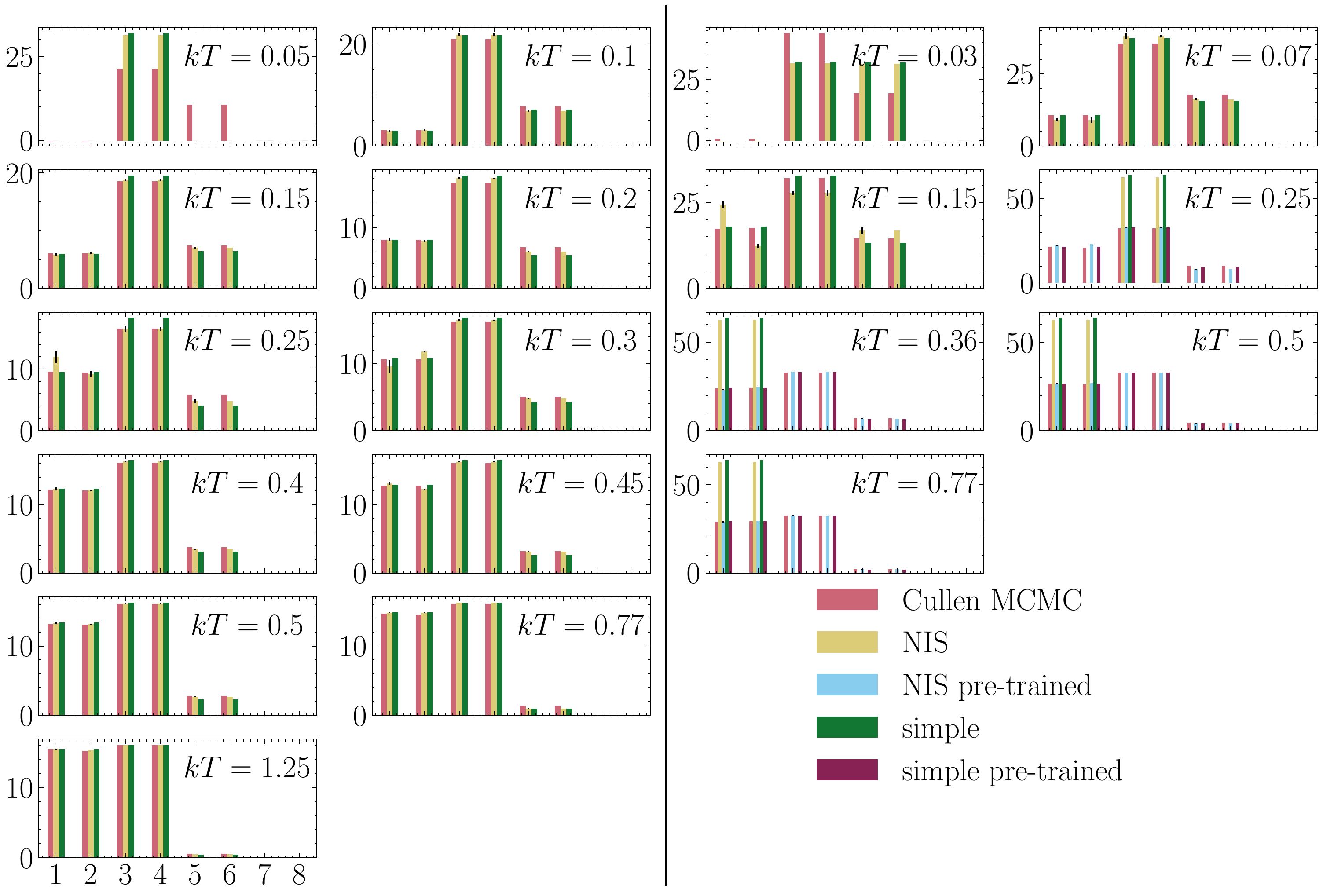}
	\caption{Average number of each plaquette $(i)$ across all samples of the Trotter approximation of an XY chain with $m=2$ at temperatures \SIlist{0.05;0.1;0.15;0.2;0.25;0.3;0.4;0.45;0.5;0.77;1.25}{} (left) and $m=4$ at temperatures \SIlist{0.03;0.07;0.15;0.25;0.36;0.5;0.77}{} (right) from a PixelCNN model and the MCMC algorithm by Landau and Cullen. The NIS estimate for each plaquette count is also given. }
	\label{fig:XY_PD}
\end{figure*}

\begin{figure*}[htpb]
	\includegraphics[width=\textwidth]{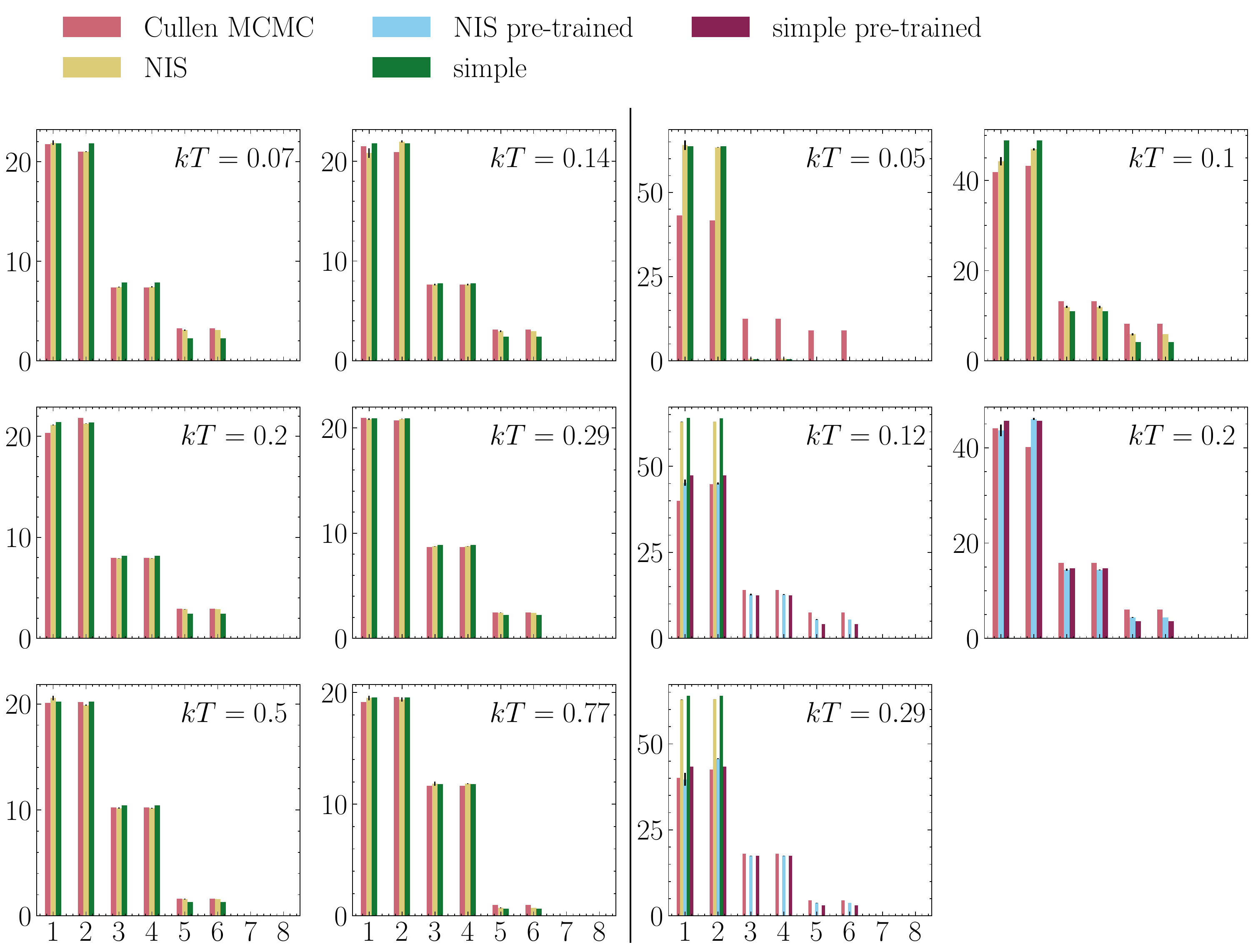}
	\caption{Average number of each plaquette $(i)$ across all samples of the Trotter approximation of an almost isotropic Heisenberg chain with $m=2$ at temperatures \SIlist{0.07;0.14;0.2;0.29;0.5;0.77}{} (left) and $m=4$ at temperatures \SIlist{0.05;0.1;0.12;0.15;0.29;0.5}{} (right) from a PixelCNN model and the MCMC algorithm by Landau and Cullen. The NIS estimate for each plaquette count is given as well.}
	\label{fig:XYZ_PD}
\end{figure*}

\Cref{fig:XY_PD} and \Cref{fig:XYZ_PD} show the plaquette distributions as obtained from different algorithms for the XY and the isotropic Heisenberg chain and the isotropic Heisenberg chain, respectively. 

\bibliography{heisenberg}
\bibliographystyle{utphys}

\end{document}